# Structural Origin of Recovered Ferroelectricity in BaTiO$_3$ Nanoparticles

H. Zhang[1], S. Liu[1], S. Ghose[2], B. Ravel[3], I. U. Idehenre[4,5], Y. A. Barnakov[4,5,6], S. A. Basun[4,5], D. R. Evans[4,*], and T. A. Tyson[1,*]

[1]Department of Physics, New Jersey Institute of Technology, Newark, NJ 07102
[2]Photon Science Division, Brookhaven National Laboratory, Upton, NY 11973
[3]National Institute of Standards and Technology, Gaithersburg, MD 20899
[4]Air Force Research Laboratory, Materials and Manufacturing Directorate, Wright-Patterson Air Force Base, OH 45433
[5]Azimuth Corporation, 4027 Colonel Glenn Highway, Beavercreek, OH 45431
[6]IPG Photonics Corporation, Southeast Technology Center Birmingham, AL 35211

## Abstract

Nanoscale BaTiO$_3$ particles (≈10 nm) prepared by ball-milling a mixture of oleic acid and heptane have been reported to have an electric polarization several times larger than that for bulk BaTiO$_3$. In this work, detailed local, intermediate, and long-range structural studies are combined with spectroscopic measurements to develop a model structure of these materials. The X-ray spectroscopic measurements reveal large Ti off-centering as the key factor producing the large spontaneous polarization in the nanoparticles. Temperature-dependent lattice parameter changes reveal the sharpening of the structural phase transitions in these BaTiO$_3$ nanoparticles compared to the pure nanoparticle systems. Sharp crystalline-type peaks in the barium oleate Raman spectra suggest that this component in the composite core-shell matrix, a product of mechanochemical synthesis, stabilizes an enhanced polar structural phase of the BaTiO$_3$ core nanoparticles.

*Corresponding Authors:  T. A Tyson, e-mail: tyson@njit.edu
D. R. Evans, email: dean.evans@afrl.af.mil

# I. Introduction

Perovskite barium titanate BaTiO$_3$ is a classical ferroelectric perovskite material known since the 1940s, and has a well-defined structure-property relationship. It undergoes several phase transitions upon cooling from high temperature including a high-symmetry paraelectric cubic to tetragonal transition at 393 K, a transition to an orthorhombic ferroelectric phase at 278 K, and a transition to a rhombohedral ferroelectric phase at 183 K. The key mechanisms of such structural transformations are based on cation displacements and oxygen octahedra rotations around different axes of the high-symmetry parent cubic phase. In the bulk system, the off-centering of Ti and O ions from the high symmetry positions in the cubic phase drives the ferroelectric state.

With reduced size, perovskite nanoparticles often possess unique electronic, optical, and magnetic properties compared to their bulk form [1]. Nanoparticles have very promising potential usage in many areas, including data storage, bio-related imaging, and targeted drug delivery [2]. Classical theoretical models predict that as the particle size is reduced, the transition temperature and ferroelectric polarization should be reduced monotonically [3], as has been demonstrated experimentally by Zhao *et al.* [3(g)]. On the other hand, it has been indicated by first-principles theoretical models that below 5 nm robust off-centering in BaTiO$_3$ nanoparticles (robust with respect to capping matrix materials) may exist in freestanding nanoparticles of BaTiO$_3$ [4]. BaTiO$_3$ has also been reported to have a linearly ordered and monodomain polarization state at nanometer dimensions, as well as room-temperature spontaneous polarization down to particles of size ∼10 nm [1(b)] in cube-shaped particles. However, the piezoelectric coefficient (d$_{33}$) in these particles is found to be ≈ 2% of the bulk value.

It was later proposed and experimentally demonstrated that ball-milling of micron-sized BaTiO$_3$ powder in oleic acid and heptane resulted in small nanoparticles (≈10 nm) with an extremely large spontaneous polarization [5(a), 7]. This "top-down" approach is explicitly suitable for ferroelectric materials, where the ferroelectric properties could be substantially modified by mechanical means: pressure and shear [3(f)]. Previous work involved doping of a liquid crystal (LC) host with ball-milled BaTiO$_3$ nanoparticles, which reported on a novel class of nematic ferroelectric LC nanocolloidal suspension [6(b)-6(d)]. Systematic electrical characterization of nanoparticles ball-milled in a mixture of heptane and oleic acid revealed up to a 5-fold enhancement of the spontaneous polarization (yielding values of 100 to 120 µC/cm$^2$) at room temperature [5(a), 5(d)]. More recent work has provided support for crystalline barium oleate being responsible for maintaining the ferroelectric properties of BaTiO$_3$ nanoparticles (i.e., spontaneous polarization of 130 µC/cm$^2$), which is formed in a mechanochemical reaction during ball-milling [6(a), 7]. In particular, in Ref. [7], it was inferred from spectroscopic studies that nano-sized BaTiO$_3$ particles (≈10 nm), and mechanochemically synthesized crystalline barium oleate formed from oleic acid in a mechanochemical reaction during ball milling, forming a core-shell type structure; it was suggested that the interface of the crystalline shell and core particle was the source of the surface stress. It is argued that the presence of such a crystalline shell plays the crucial role in the maintenance and even an enhancement of ferroelectricity on the nano-scale. It is suggested that the core-shell structure is created via the formation of specific bonding between surface atoms of the nanoparticle and carboxylate ions of organic ligands at the interface, which results in a lattice mismatch between the crystalline shell and core [7]. Note, the core-shell systems described in Refs. [6(a), 7] are used for the work described in this current paper; they are also very similar in nature to the materials that were investigated in Refs. [3(f), 5(a), 5(c), 5(d), 6(b), 6(c), 6(d), 6(e)]. We note that no direct structural measurement has yet been given in support of the existence of the core-shell model.

Although it is known that these nanoscale materials exhibit an electric polarization significantly exceeding bulk values [5(a), 5(d), 7], no direct structural information (such as bond length changes, distortions, and structural changes with temperature) is available for this system. This current work aims to provide a detailed understanding of the atomic-level structure of the final nanoparticles prepared using the ball-milling process (i.e., the mechanochemical synthesis process). This will help enable the development of a general approach for the stabilization and enhancement of the ferroelectric state in bulk systems when the size is reduced to the nanoscale. The primary results of the analysis in this work is that a significant Ti off-centering is present concomitant with the appearance of polar phonon modes, and a ratio of $BaTiO_3$ to barium oleate of $\approx$1:1 is found suggesting a coating of the $BaTiO_3$ particles by the crystalline barium oleate.

## II. Sample Preparation, Experimental Approach, and Modeling Methods

The 10 nm nanoparticles were prepared by mechanical ball-milling commercial bulk $BaTiO_3$ for 25 hours in heptane and oleic acid using a Retsch[1] planetary high energy ball-mill (PM 200) with zirconia crucibles and 2 mm zirconia beads. The ratio of $BaTiO_3$, oleic acid, and heptane used was 1:1:20 by weight [5(c)]. The final particle size was determined by transmission electron microscopy (TEM) as well as powder X-ray Diffraction (XRD) refinement. In this work, the particle diameter from ball-milling was determined to be 9(2) nm (Scherrer method) [8] (see supplementary document Fig. S1 and Table S1). The primary samples used in the measurements were washed with anhydrous ethanol to remove excess unreacted freestanding oleic acid/carboxylate from the colloidal suspension. The supplementary document presents data for both washed and as-prepared samples. Unmilled $BaTiO_3$ nanoparticles with 700 nm and 50 nm diameters (actual size is 70(3) nm (Scherrer method, Table S1)) were purchased commercially from Alfa Aesar and are used as reference samples. The unmilled 700 nm $BaTiO_3$ is considered to be a bulk standard. Commercial barium oleate (Alfa Aesar) was milled and used as a reference sample.

Laboratory XRD measurements were combined with synchrotron-based diffraction and spectroscopy experiments to determine the structural details of this system. More details on the specific measurement systems and modeling used, as well as detailed Raman measurements and x-ray diffraction modeling, can be found in the Supplementary Document.

---

[1] Certain commercial equipment, instruments, or materials are identified in this paper to foster understanding. Such identification does not imply recommendation or endorsement by the National Institute of Standards and Technology, nor does it imply that the materials or equipment identified are necessarily the best available for the purpose.

# III. Results and Discussion

## III(a). Raman Measurements

The results from Raman measurements are shown in Fig. 1 and in Figs. S2 and S3 (supplementary document). Figure 1 shows the room temperature Raman spectra for the unmilled 700 nm (bulk) and 50 nm reference samples and the milled 10 nm BaTiO$_3$. Also shown in Fig. 1 are data for barium oleate; for comparison, oleic acid data is presented in Fig. S2 (taken from (Ref. [9]). Note that the oleic acid starting material is converted to a metal carboxylate (i.e., barium oleate) during the ball-milling process via mechanochemical synthesis [6(a), 7].

The arrow in Fig. 1(a) reveals that for both the 50 and 10 nm samples, there is an additional peak near 190 cm$^{-1}$, which is not present in the 700 nm (bulk) sample indicating a reduction in symmetry. This feature is known to become enhanced in the rhombohedral low-temperature phase in bulk samples (below ≈183 K) with micron-scale grains, but it is evident in these nanoparticles at room temperature. This peak corresponds to an A1 (TO) mode in the tetragonal system as seen by Marssi *et al.* in Ref. [10], which is significantly sharpened in the 10 nm sample. These modes are characteristic of a polar phase in perovskite systems and their strong enhancement indicates a profound symmetry reduction in the 10 nm sample consistent with polar type space groups.

It can also be seen that there are additional peaks in the spectra of the 10 nm sample, and a strong similarity in peaks in the barium oleate sample above ≈ 800 cm$^{-1}$ (see Figs. 1(b), S2, and S3), which are not present in the 50 nm and 700 nm samples. Note, that oleic acid also shows similarities to this (see Figs. S2 and S3), but there are significant deviations over the range of 800 cm$^{-1}$-1200 cm$^{-1}$. The high intensity of these features in the 10 nm sample indicates the presence of a significant contribution of organic components. A complete set of measurements of all samples between 50 cm$^{-1}$ and 3400 cm$^{-1}$ is shown in the supplementary document (Fig. S2 and S3).

The assignment of the aforementioned spectral features is as follows: the high-frequency peaks are due to light atoms and hence the spectra at these frequencies do not provide information of bonding to heavy atoms (such as the Ti or Ba atoms in the starting material); in systems containing carbon chains, the C-C single bonds generate modes near 1000 cm$^{-1}$, while double-bonded carbon produces modes near 1600 cm$^{-1}$ [11]; peaks near ≈ 3000 cm$^{-1}$ correspond to C-H bonds. Incidentally, the set of features that shows a significant difference between the samples is over the range of 1000 cm$^{-1}$ - 1150 cm$^{-1}$ (Fig. S2(b)); in this case, the barium oleate sample shows the strongest resemblance to the 10 nm nanoparticles. This suggests that barium oleate may be playing a more significant role in the nanocolloids than the oleic acid; this is in agreement with the Fourier-transform infrared spectroscopy (FTIR) results in Refs. [6(a)] and [7], which demonstrates nearly all of the oleic acid is converted to a metal carboxylate (i.e., barium oleate) during the milling process via mechanochemical synthesis. For this reason, the discussion below is centered around barium oleate as the primary organic component in the ferroelectric nanocolloid (milled 10 nm sample).

A comparison of the peaks above 800 cm$^{-1}$ for the 10 nm nanoparticle and barium oleate samples reveals that all features are present (Fig. 1(b)) for both cases; these features are absent in the unmilled 50 nm and 700 nm reference samples. The sharpening of the peaks in the 10 nm sample is due to a significantly higher degree of structural order compared to crystalline barium oleate. The results point to a complex BaTiO$_3$/barium oleate composite mixture (possibly a core-shell configuration as suggested in Ref. [7] or nanoparticles embedded in a barium oleate host). Details of the structure were found by systematic x-ray absorption and x-ray pair distribution function measurements, which are preferentially sensitive to heavy atoms (see below).

## III(b). X-Ray Absorption Measurements and Simulations (Local Structure about Ti and Ba)

Figure 2 (a) shows the Ti K-edge x-ray absorption near-edge structure (XANES) for the 700 and 50 nm reference samples compared to the milled 10 nm sample. (Spectra presented in Fig. 2(a) and Fig. 2(b) are the result of averaging two to three consecutively measured scans. The statistical variations in the scans are the level of the line thicknesses since the samples are concentrated. The R-space curves in Fig.

2(c) and 2(d) were extracted from the corresponding averaged spectra). The 700 and 50 nm XANES spectra exhibit sharp peaks in the region near 4960 eV (near-threshold (see Fig 2(a) inset)) as expected for bulk-like systems with high atomic order, while the 10 nm sample is broadened due to a reduction in long-range structural coherence relative to the bulk sample. Focusing on Fig. 2(a), it is seen that the pre-edge peak near 4968 eV (Feature A) of the 10 nm sample is broadened (and more intense) compared to both the 700 nm (bulk) and 50 nm samples. This feature is well-known to be related to the degree of off-centering of the Ti sites (relative to the cubic structure) in ATiO$_3$ systems concomitant with a polar phase (see Refs. [12, 13] and references therein). These multiple pre-edge features near $\approx$ 4970 eV correspond to a transition from Ti 1s $\rightarrow$ Ti 4p hybridized with Ti 3d and O 2p (analogous to the well-studied perovskite manganites [14]). Feature A in the spectrum corresponds to a transition to the $e_g$ band, and the unlabeled lower energy shoulder (left side of Feature A) is related to the $t_{2g}$ band. The $e_g$ peak in the XANES spectra of ATiO$_3$ systems is sensitive to local distortions (i.e., coordination of the Ti sites) and is found to have an integrated intensity proportional to the mean square displacement of the Ti atom off the ideal (cubic) site [13(b)]. Hence the higher intensity of this peak in the 10 nm sample reveal larger off-centering than in the reference samples. This is revealed more clearly by the simulations below.

XANES simulations for the pre-edge region are shown for specific structural changes in Fig. 2(b) (See Ref. [12] and supplementary document for structure configuration details and computational methods). The simulations yield the XANES pre-edge spectra in the observed BaTiO$_3$ structure with either a 0.12 Å Ti z-displacement or with random displacements in x, y, and z directions (with a 3-D random displacement of all atoms giving average amplitudes (W) of 0.06 Å or 0.15 Å); these displacements are relative to the room temperature *orthorhombic* atomic positions in the lattice. The spectra of cubic BaTiO$_3$ with no distortion and that for *cubic* BaTiO$_3$ with a 0.15 Å average atomic distortion are also given as a reference. Note the weakest main peak amplitude (Feature A) is obtained for the undistorted cubic phase, while the peak amplitude is significantly enhanced for both the Ti off-centering in the "normal" orthorhombic phase and random (incoherent) disorder characteristic of amorphous phases (see amorphous example in Ref.

[13(a)]). It is noted that for the case of Ti displacements ($\Delta z$), Feature A is enhanced while Feature B is suppressed. The sample is not cubic since there are polar Raman modes. We also observe that the Raman data reveal no significant peak broadening in the nanoscale material, there is also no evidence of a cubic structure found below in standard XRD (x-ray diffraction) or PDF (pair distribution function) measurements. The milled 10 nm sample is not amorphous. (As seen in the PDF measurements below the peaks in the radial distribution function beyond the first shell are not suppressed). Hence, the enhanced peak area of the main pre-edge feature, A, seen in the 10 nm $BaTiO_3$ sample compared to bulk $BaTiO_3$ is due to Ti off-centering in a crystalline phase characteristic of the polar ferroelectric state. (It should be noted that the 50 nm reference sample has a weakly enhanced area relative to the 700 nm bulk-like sample).

The extended x-ray absorption fine structure data are consistent with the XANES results. Figure 2(c) shows the Fourier transform of the fine structure over the k-range ($2.00 < k < 7.70$ Å$^{-1}$) for a qualitative comparison. This structure function has peaks that correspond to the atomic shells about the average Ti site. The results at the Ti K-edge indicate a strong Ti-O first shell coordination, but there is a significant reduction of structural order for the second shell Ti-Ba and higher shells in the 10 nm sample (the Ti-Ba and Ti-Ti peaks are suppressed in the 10 nm sample). This reduction is due to structural disorder or lowering of symmetry about the Ti sites (variation in nearest neighbor Ti-Ba and Ti-Ti distances) leading to destructive interference of the higher shell scattering signals. Ti-Ba and Ti-Ti signals with different bond distances interfere destructively and suppress the peak amplitudes. This suggests large structural disorder or highly reduced symmetry in the nanoparticles. Examining Figs. 2(a) and 2(c), it is seen that from the perspective of the Ti sites that new chemical phases containing Ti are not evident. The primary change in the 10 nm sample is the reduction in structural correlation between Ti and higher neighbors beyond O. Combined with the Ti XANES results, the characteristic structural feature of the milled 10 nm sample with respect to bulk (orthorhombic) $BaTiO_3$ is the large Ti off-centering.

For a qualitative comparison, Fig. 2(d) shows the Fourier transform of the fine structure for the Ba L3-edge over the k-range $1.98$ Å$^{-1} < k < 9.43$ Å$^{-1}$. This structural function has peaks that correspond to the atomic shells about the average Ba site. The 700 nm and 50 nm reference samples show similar structure relative to Ba sites (as in the case of the Ti site), however, the 10 nm sample has a first shell peak shifted to lower R values. Also shown in Fig. 2(d) are data for barium oleate. We note that the shift and position of the Ba-O first peak in the 10 nm sample are consistent with that of barium oleate (black curve). The results suggest that a significant component of barium oleate is present in the sample, consistent with the Raman results (above) and the FTIR data presented in Ref. [6(a)]. The diffraction measurements in the following paragraph provide strong support for this assertion.

## III(c). Wide Angle X-Ray Diffraction (Long-Range Structure)

Figure 3 shows the results from laboratory XRD measurements, including data for the 10 nm, 50 nm, and barium oleate samples. Barium oleate has peaks primarily below 30º (2θ). Further details can be seen by expanding the figure, see Fig. 3(b), where the barium oleate sample is compared with the 10 nm sample; it becomes evident that there is a structure present that is similar to barium oleate although some of the peak intensities do not fully match. This suggests that a derivative of barium oleate (with minor changes in atomic position) is present and possibly coating the $BaTiO_3$ particles or hosting them in a matrix. Some possible modifications may be due to the formation of chelate or bidentate structures as suggested in Ref. [7].

It is important to note that the sharp Bragg peaks in Figs. 3(a) and 3(b) for the 10 nm sample $BaTiO_3$ sample indicate that both the milled 10 nm $BaTiO_3$ and the milled "barium oleate" type components are highly crystalline. No broad maxima, typical of amorphous components, are seen in the data. It should also be noted that the peak widths in barium oleate and the 10 nm $BaTiO_3$ samples are quite similar indicating a high degree of crystallinity in the nanoscale material with respect to the heavy atom sites which dominate the x-ray diffraction pattern. To further elucidate the nature of the structure of 10 nm $BaTiO_3$ samples for both local ($< \approx 5$ Å) and intermediate ranges x-ray PDF measurements were conducted.

## III(d). Pair Distribution Function Measurements and Modeling

Total scattering x-ray PDF measurements (which includes both the Bragg scattering and diffuse scattering due to local distorions) were conducted to investigate the details of the structure. The atomic pair distribution functions G(r) are displayed in Fig. 4. Note that G(r) is the reduced pair distribution function, which oscillates about zero and is obtained directly from the scattering data, S(Q). $G(r) = \frac{2}{\pi}\int_0^\infty Q[S(Q)-1]\sin(Qr)dQ$ is related directly to the standard pair distribution function defined as $G(r) = 4\pi r(\rho(r) - \rho_0)$, where $\rho(r)$ is the atomic number density and $\rho_0$ is the average atomic number density. In Fig. 4(a), it is seen that for a radius, r, greater than ≈10 Å the shape of the pair distribution function for the 700, 50, and 10 nm samples are fairly similar. Although for the 10 nm sample, peaks and troughs are broader due to the particle size effect. Below ≈10 Å there is a significant difference between the milled 10 nm sample and the unmilled 50 and 700 nm samples. As in the case of the high-frequency Raman data, laboratory XRD data, and Ba $L_3$-edge x-ray absorption fine structure (XAFS) data (Figs. 1-3), the difference between the 10 nm sample and the $BaTiO_3$ model indicates the existence of another compound with a significant organic component similar to the barium oleate reference sample. Note that the residual (i.e., the signature of an additional species that is observed when the model fit results are subtracted from the 10 nm $BaTiO_3$ data) has sharp and well-defined peaks below 10 Å indicating that both the $BaTiO_3$ and the extra component (second phase) are highly ordered structurally. Barium oleate was used as a possible second phase by comparison to the residual of the fit of the data to the $BaTiO_3$ model. PDF measurements were conducted under the same conditions as for the $BaTiO_3$ samples.

To obtain a better understanding, structural refinements with models were conducted [15]. In order to model the system with a limited set of free parameters, a tetragonal model was utilized. We note the real system may be orthorhombic. The main point is that the 10 nm system was found to be non-cubic (a and c are not equal).

Experimental PDF patterns are shown in Fig. 4(a). Model fits of the region between 10 Å and 20 Å were conducted for a tetragonal model of $BaTiO_3$. For the 10 nm sample, the refined structural parameters were then used to calculate the G(r) for r ≤10 Å; the result is shown in Fig. 4(b). In this panel the red curve is the experimental PDF, the blue curve is the model, and the green curve is the difference between the data and the model (residual). It is clear that the residual profile (data minus the fit for the 10 nm sample) matches well with the G(r) of the barium oleate (black curve).

To quantify the amount of barium oleate relative to $BaTiO_3$, a barium acetate phase structural model (which possesses similar short-range structure to barium oleate), was included in the structural refinement. Hence the model included $BaTiO_3$ and barium acetate (model for barium oleate). The result is shown in Fig. 4(c). As shown in Fig. 4(b), the residual is the difference between the data and the model. The extra features in the low r-range now can all fit well, except for the full amplitude of the nearest neighbor Ba-Ba ≈ 4 Å. The ratio of the two components ($BaTiO_3$ and the barium acetate model) acquired from refinement is 1:0.76, indicating a very significant amount of barium oleate existing in the $BaTiO_3$ 10 nm ball-milled sample. The exact form of barium oleate is not known. However, as seen from the very sharp peaks in the barium oleate component from Raman measurements, it is not freestanding barium oleate, but may be bonded to the $BaTiO_3$ and its structure may be modified. The results suggest that the $BaTiO_3$ particles are coated by barium oleate (consistent with optical data in Ref. [7]) with ≈ 1:1 $BaTiO_3$ More accurately, the Ba site:Ba site (i.e., barium in the core $BaTiO_3$ and in barium oleate sites) ratio in both components is ≈ 1:1 (by volume or number of Ba).

To understand the phase transitions in the $BaTiO_3$ core-shell samples, the lattice parameters derived from temperature-dependent PDF refinements were determined between ≈ 100 K and ≈ 500 K. Nonlinear variations in the lattice parameters (or their ratios) are expected at the cubic to tetragonal transition near 393 K, the tetragonal to orthorhombic transition near 278 K, and the orthorhombic to rhombohedral transition near 183 K (see the structural transitions in bulk $BaTiO_3$ in Ref. [16]). In Fig. 5, the temperature dependence of the c/a lattice parameter ratio is given. For the (unmilled) 700 nm reference sample (bulk), inflection points (indicated by arrows) are seen at the appropriate positions for the phase transition temperatures. For the (unmilled) 50 nm reference sample, the transitions are broadened significantly and are indistinguishable. In particular, the low-temperature tetragonal to orthorhombic and orthorhombic to rhombohedral transitions are merged. This broadening of the transition with reduced particle size was observed for freestanding $BaTiO_3$ particles for sizes down to 28 nm in previous work [17], where the broadening was assigned to loss of structural coherence of the local distortions. In contrast to the 50 nm sample, the transitions in the milled 10 nm sample are readily visible. Also, a c/a ratio with significant deviations from unity (right scale of Fig. 5) exists for the temperate range ≈ 150 K to ≈ 500 K. These results suggest that the 10 nm sample possesses a stong bulk-like behavior including stabilization of the polar ferroelectric phase.

## IV. Summary

$BaTiO_3$ nanoparticles (≈10 nm) prepared by ball-milling in oleic acid are studied together with unmilled 50 nm and 700 nm (bulk) particles as a reference. Raman spectroscopy, laboratory XRD, synchrotron-based XAFS, and PDF analysis have been conducted to understand the structural mechanism behind the unusually large electric polarization in the nanoparticles. Compared to freestanding barium oleate, the corresponding component in the 10 nm sample matrix exhibits sharp Raman peaks consistent with a highly crystalline form, which is presumed to be crystalline organic coating (shell) around the $BaTiO_3$ core with a coating to core ratio of ≈ 1:1. Local structural measurements show large Ti off-centering in the milled 10 nm samples (enhanced beyond orthorhombic (bulk) $BaTiO_3$). Moreover, bulk-like sharp structural transitions are observed in the milled 10 nm nanoparticles, in contrast to the 50 nm unmilled nanoparticles. A complex crystalline $BaTiO_3$/Ba Oleate type composite material exists, which stabilizes a high a/c lattice asymmetry and in turn a strong ferroelectric effect.

Considering this nanoparticle system to be core-shell in configuration, the recovery of bulk-like behavior is attributed to stress provided by the barium oleate outer shell, i.e., a lattice mismatch between the crystalline organic shell and inorganic core, which is responsible for sustaining and enhancing the ferroelectric effect (spontaneous polarization) in these nanoparticles. Surface stress is demonstrated experimentally by Cook *et al.* in Ref. [3(e)]. Surface stress enhancing ferroelectric properties is supported theoretically by Morozovska *et al.* (Ref. [3(d)]) and Ederer and Spaldin (Ref. [18(b)]). Both in-plane compression and in-plane tension are known to enhance ferroelectricity in $BaTiO_3$ thin films [18]. The same mechanism is expected in this case. We note that studies of $CoFe_2O_4$ nanoparticles reveal that oleic acid molecules bond to metal sites on the surface of the particles and produce surface strain resulting in enhanced coercivity [19]. In the case of $BaTiO_3$, the strain enhances the Ti off-centering resulting in an enhanced spontaneous polarization.

## V. Acknowledgments

This work is supported in part by the U.S. Air Force under Grant FA8650-16-D-54 and the applied physics and materials science programs of the physics department at NJIT. Synchrotron X-ray absorption fine structure experiments were performed at beamline 6 BM (BMM) at the National Synchrotron Light Source II (NSLS 2). PDF measurements were conducted at beamline XPD2 at NSLS2 and beamline X17A at NSLS, a U.S. Department of Energy (DOE) Office of Science User Facility operated for the DOE Office of Science by Brookhaven National Laboratory under Contract No. DE-SC0012704.

**Figure Captions**

**Fig. 1.** Room temperature Raman measurements for 700 nm bulk (blue), 50 nm (green), and milled 10 nm (red) BaTiO$_3$ samples are compared with barium oleate (black open triangles). Spectra are shown for the range (a) 150 nm to 750 cm$^{-1}$ and (b) the higher energy region 800 nm to 1200 cm$^{-1}$. Data for the range 50 cm$^{-1}$ to 3400 cm$^{-1}$ are given in supplementary document Figs. S2 and S3. Data for the 10 nm samples matches barium oleate over the spectral range 800 cm$^{-1}$ - 3400 cm$^{-1}$ (Figs. 1b, S2, and S3). The peaks of the 10 nm sample are significantly sharper, indicating a high degree of crystalline order. The arrow in Fig. 1(a) identifies an additional peak near 190 cm$^{-1}$ for the 50 nm and 10 nm samples indicating a reduction in symmetry; this peak is not present in the 700 nm (bulk) sample. In (b) the 10 nm nanoparticle and barium oleate samples are shown for comparison.

**Fig. 2.** (a) Curves for Ti K-edge XANES spectra for 10 nm BaTiO$_3$ (red), 50 nm BaTiO$_3$ (green), and 700 nm BaTiO$_3$ (blue). (b) Simulated XANES pre-edge spectra for bulk BaTiO$_3$ (orthorhombic) called "Normal", BaTiO$_3$ (orthorhombic) with 0.12 Å z-displacement, and BaTiO$_3$ (orthorhombic) with random displacements of all atoms with 0.06 Å and 0.15 Å average values. The spectra of BaTiO$_3$ with no distortion (cubic) and with 0.15 Å average atomic distortion to cubic BaTiO$_3$ are also given. (c) Ti K-edge structure function for 10 nm (red), 50 nm (green), and 700 nm BaTiO$_3$ (blue) samples. Distances are relative to the average titanium atom position. (d) The corresponding Ba L$_3$ structure functions for the same samples as in part (c). The blue, green, red, and black curves correspond to 700 nm, 50 nn, 10 nm, and barium oleate samples, respectively. Distances are relative to the average barium atom position. Note that Fig. 2(c) utilizes the same x-axis scale and range used in Fig 2(d). In both figures ((c) and (d)), note that $\chi(k)$ is dimensionless while the wavevector k has units of Å$^{-1}$.

**Fig. 3.** (a) XRD measurements of the 10 nm (red) and 50 nm (green) samples of BaTiO$_3$, and barium oleate (black). (b) Expanded region between 15° to 30° in 2Θ for the 10 nm and barium oleate samples.

**Fig. 4.** (a) Experimental PDF G(r) curves for the 700 nm (blue), 50 nm (green), and 10 nm (red) BaTiO$_3$. The inset is the expanded region between 10 to 20 Å. (b) Experimental PDF curve (red), refinement model (blue), and the residual (olive) for 10 nm BaTiO$_3$. The experimental PDF curve of barium oleate (black) is shown for comparison. Note the similarity. (c) An additional barium acetate phase was added to the region between 1.5 and 10 Å to model the short-range features of the PDF.

**Fig. 5.** c/a ratio of the 700 nm (blue), 50 nm (green), and 10 nm (red) BaTiO$_3$ core-shell nanoparticles. The left y-axis is for the 700 nm and 50 nm reference samples, and the right y-axis for the 10 nm milled sample.

**Fig. 1.** Zhang *et al.*

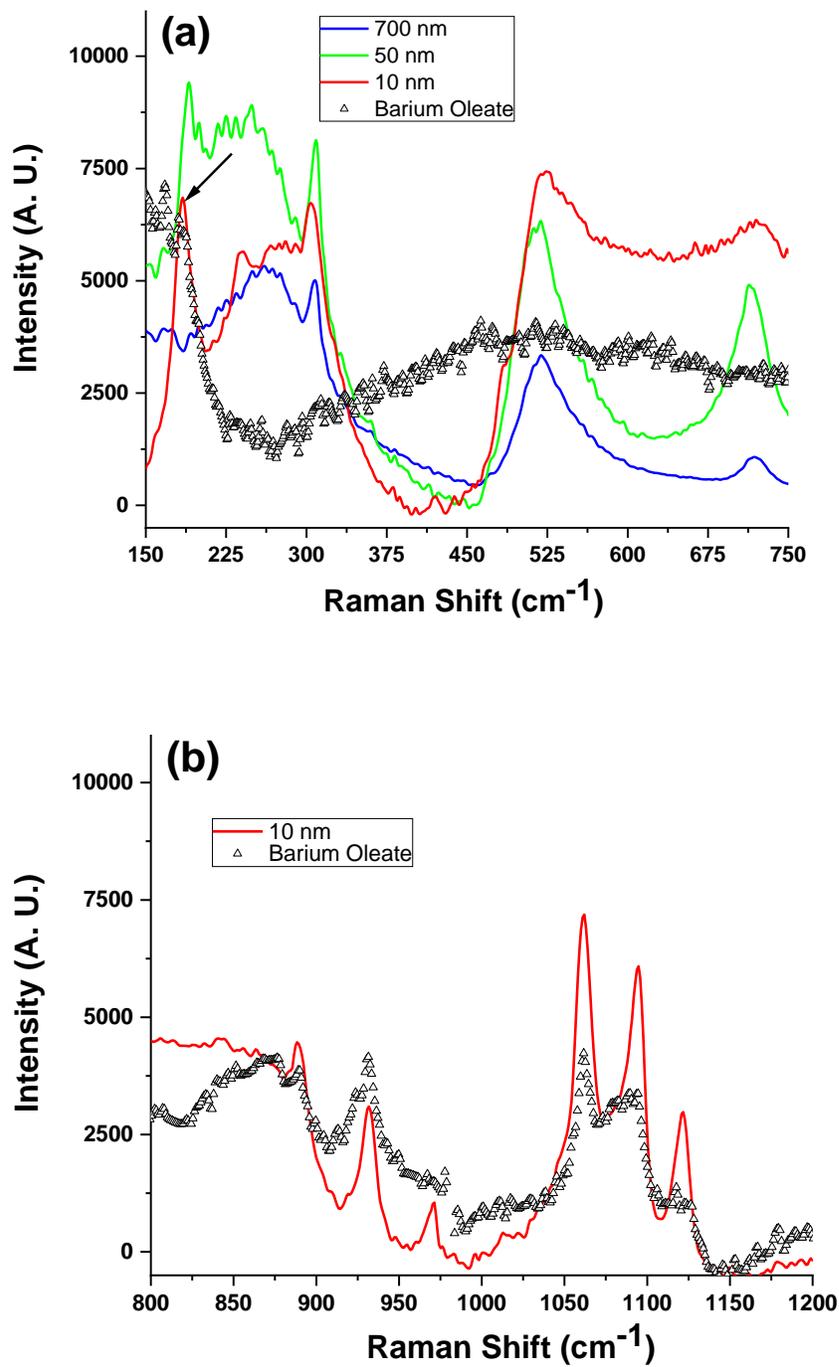

**Fig. 2.** Zhang *et al.*

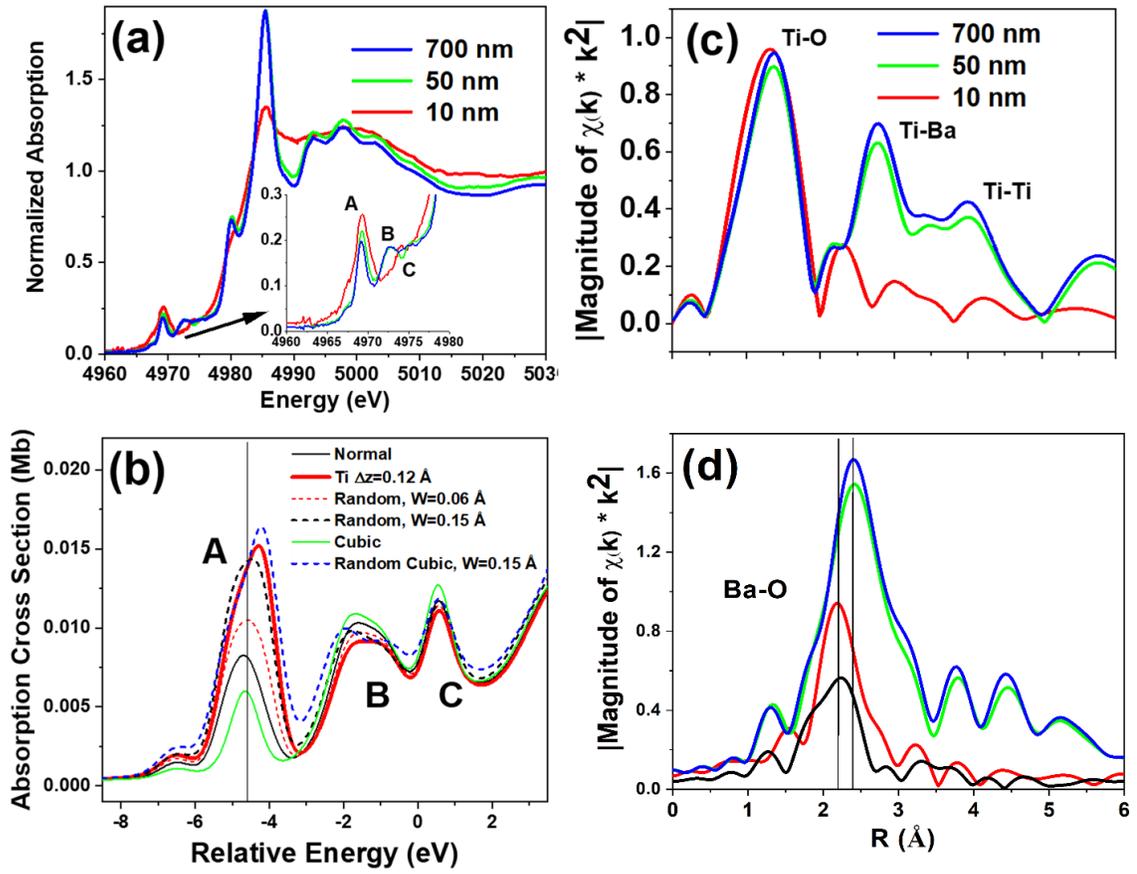

**Fig. 3.** Zhang *et al.*

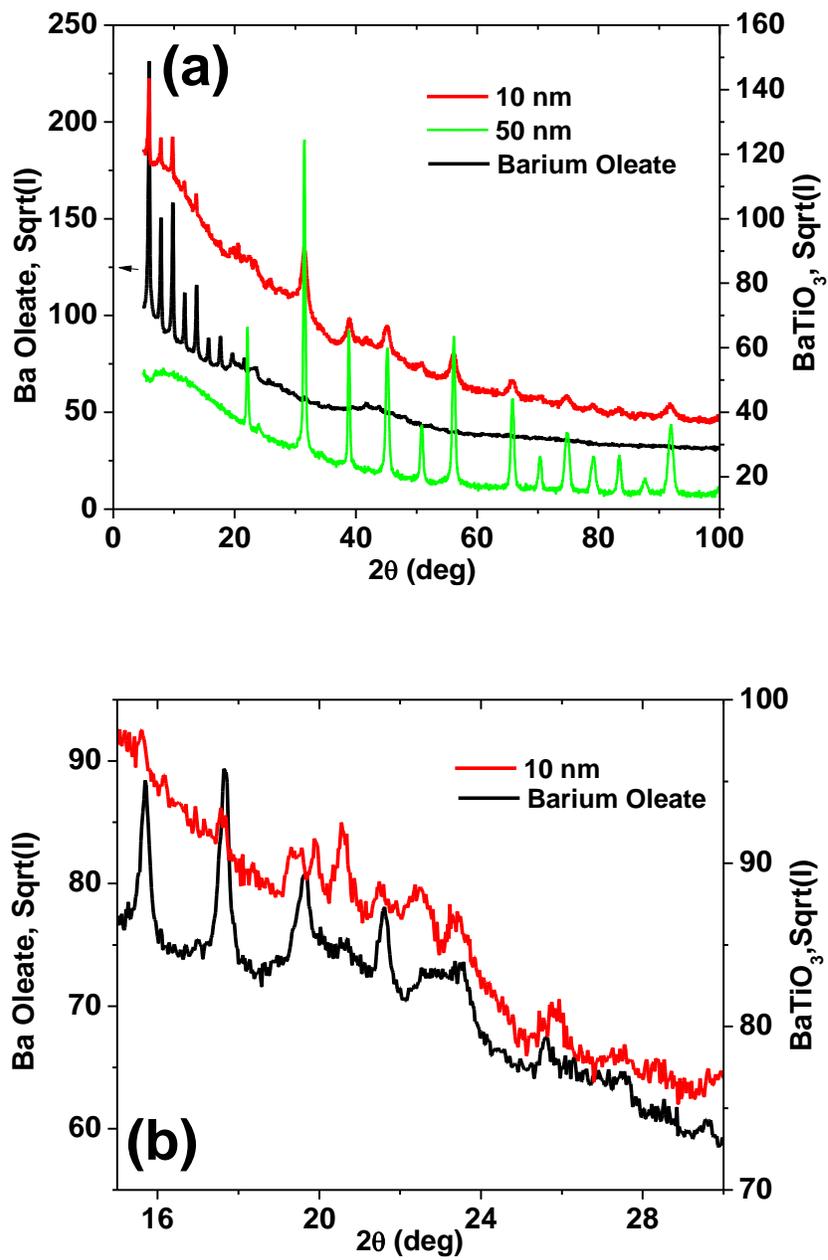

**Fig. 4.** Zhang *et al*

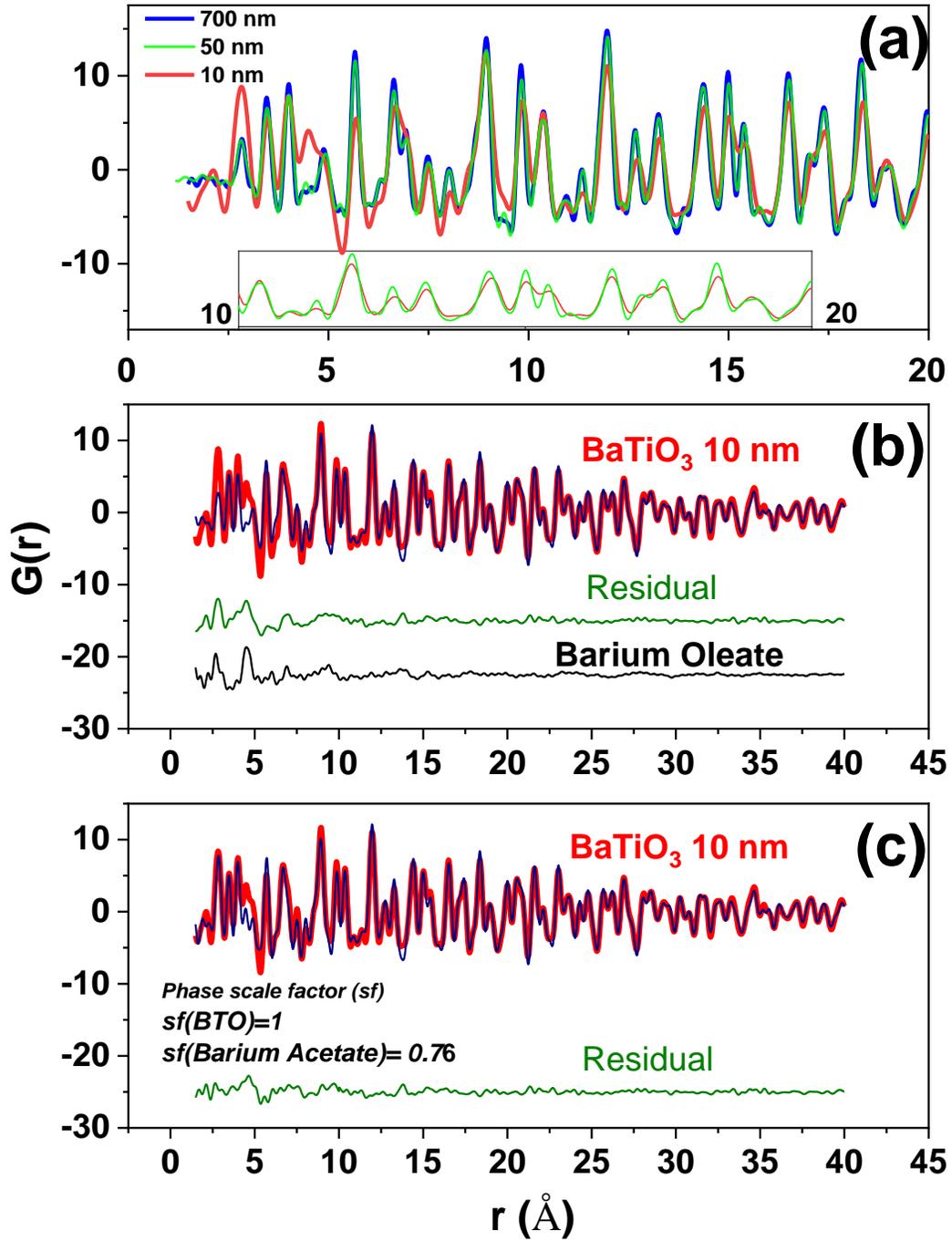

**Fig. 5.** Zhang *et al.*

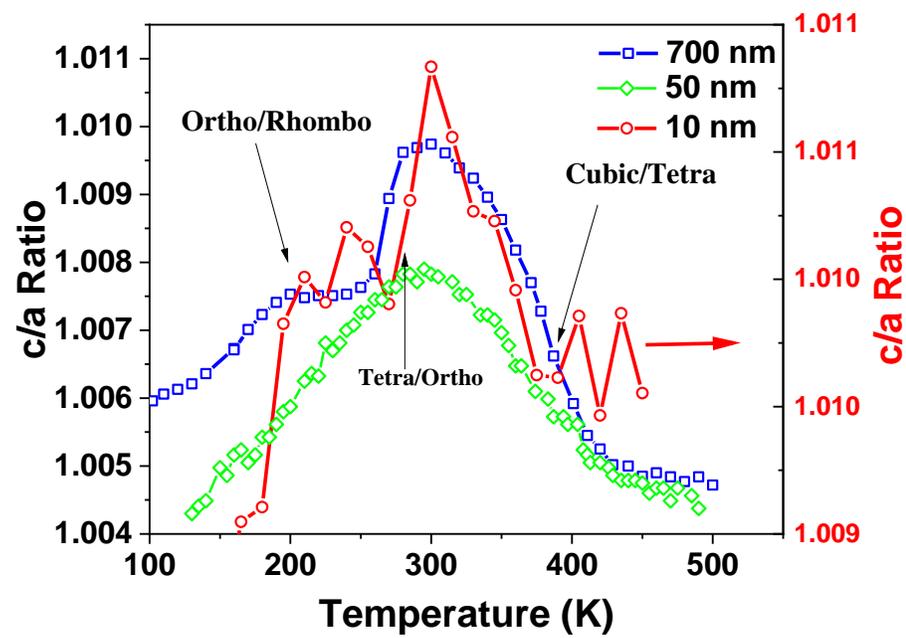

# Structural Origin of Recovered Ferroelectricity in BaTiO$_3$ Nanoparticles
## (Supplementary Document)

H. Zhang[1], S. Liu[1], S. Ghose[2], B. Ravel[3], I. U. Idehenre[4,5], Y. A. Barnakov[4,6], S. A. Basun[4,6], D. R. Evans[4,*], and T. A. Tyson[1,*]

[1]Department of Physics, New Jersey Institute of Technology, Newark, NJ 07102
[2]Photon Science Division, Brookhaven National Laboratory, Upton, NY 11973
[3]National Institute of Standards and Technology, Gaithersburg, MD 20899
[4]Air Force Research Laboratory, Materials and Manufacturing Directorate, Wright-Patterson Air Force Base, Dayton, OH 45433
[5]Department of Electro-Optics and Photonics, University of Dayton, Dayton, OH 45469
[6]Azimuth Corporation, 4027 Colonel Glenn Highway, Beavercreek, OH 45431

*Corresponding Authors: T. A Tyson, e-mail: tyson@njit.edu
D. R. Evans, email: dean.evans@afrl.af.mil

Laboratory Raman and x-ray diffraction measurements were conducted at the Otto H. York Center for Environmental Engineering and Science at NJIT. The Raman spectra were measured with a ThermoFisher DXR Raman Microscope using a 532 nm laser and 10x objective. The barium oleate measurements were conducted with a 50x objective using a 780 nm laser to suppress the fluorescence contribution to the signal. Laboratory x-ray diffraction studies were conducted using a Philips Empyrean x-ray diffractometer (Fig. 4 and Table S1) with a copper x-ray tube. The x-ray absorption measurements with 10 nm as-prepared, 10 nm washed, 50 nm, and 700 nm (bulk) $BaTiO_3$ were conducted at beamline 6-BM (BMM) at NSLS2 in Brookhaven National Laboratory (BNL). The PDF (Pair Distribution Function) analysis with 10 nm ethanol-washed $BaTiO_3$ and barium oleate samples utilized data collected using beamline 28-ID-2 (X-ray Powder Diffraction, XPD) at NSLS2 (National Synchrotron Light Source), at Brookhaven National Laboratory with a wavelength of 0.2366 Å. The PDF analysis with 50 nm and 700 nm $BaTiO_3$ reference samples are based on measurements from beamline X17A at NSLS, BNL. XAFS (X-ray Absorption Fine Structure) and PDF data acquisition and reduction follow the methods used in Ref. [1] for nanoscale $SrTiO_3$. Modeling of XANES (X-ray Absorption Near-Edge Spectra) data (Fig. 3(b)) was conducted using the RELXAS full multiple scattering code [2] as done in Ref. [1]. For the fits in Fig. S1, the Topas refinement package was used for data analysis [3]. Note, "washed" samples were cleaned with anhydrous ethanol to remove excess unreacted freestanding oleic acid/carboxylate from the colloidal suspension.

With respect to errors and uncertainties, we note the major results presented in Fig. 1 to Fig. 3 reveal systematic trends in phonon peaks, amplitudes, and trends in the coordination shell amplitudes and relative positions (distances) and Bragg peak shapes and positions. The primary quantitative results correspond to the size of the nanoparticles and the BTO to barium oleate ratio. We utilized the Scherrer model fits to determine the average BTO nanoparticle size to be 9(2) nm. Raman spectra were measured to determine the additional component present in the 10 samples and an appropriate model was utilized to account for this phase in the PDF fits. We conservatively estimate the barium oleate to BaTiO3 ratio to be 1(2):1, having fixed the BTO level to 1. Figure S0(b) reveals the similarity between the BTO fit residual and barium oleate data. These results are complemented by the same comparison in the Raman measurements (Fig. 1(b)). The additional component in the "10nm BTO" system is indeed a barium oleate-like structure.

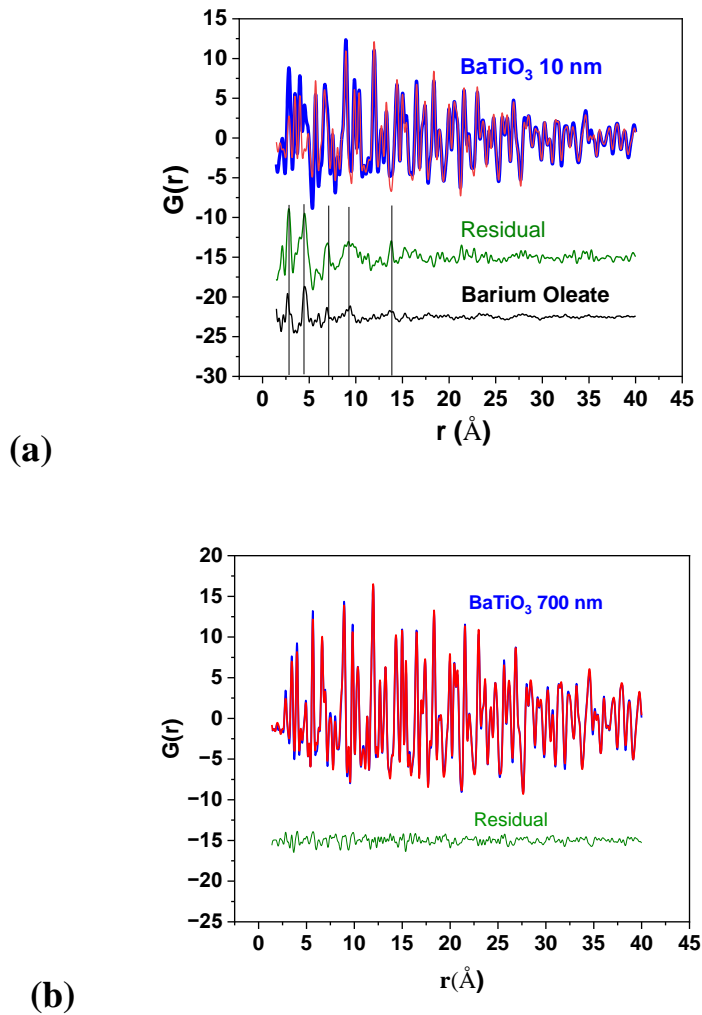

**Fig. S0**. (a) Expanded version of Fig. 4(c) with vertical lines indicating the common peaks in the fit residual and the barium oleate model system. For the upper set of curves blue corresponds to experimental data and red corresponds to the model pure BTO fit. (b) Fit of the same BTO model to the 700 nm sample data showing no strong peaks in the residual. Note also that the residual amplitude is approximately the same over the entire r-range. Note that the y-axis change is the same in (a) and (b).

**Table S1. Extracted BaTiO$_3$ Particle Diameters from Refinement of Lab XRD data***

| Sample Type | Extracted Diameter (nm) |
| --- | --- |
| 50 nm | 70 ± 3 |
| 10 nm washed | 9 ± 2 |

*The 700 nm sample was used as an infinite size (bulk) reference.

## (a) 700 nm-Bulk

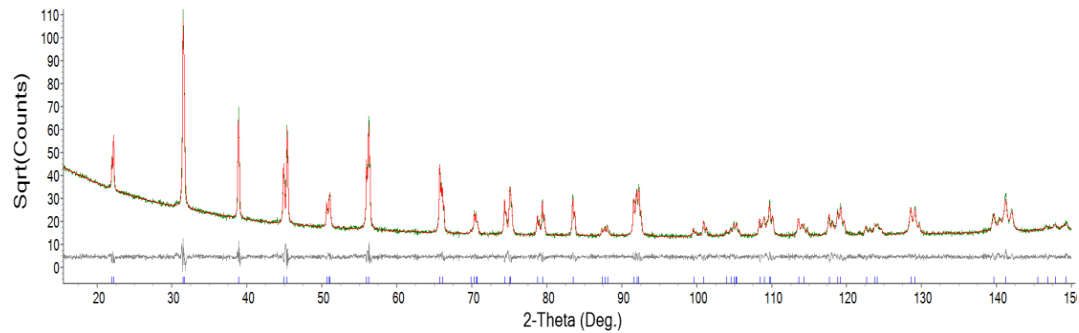

## (b) 50 nm

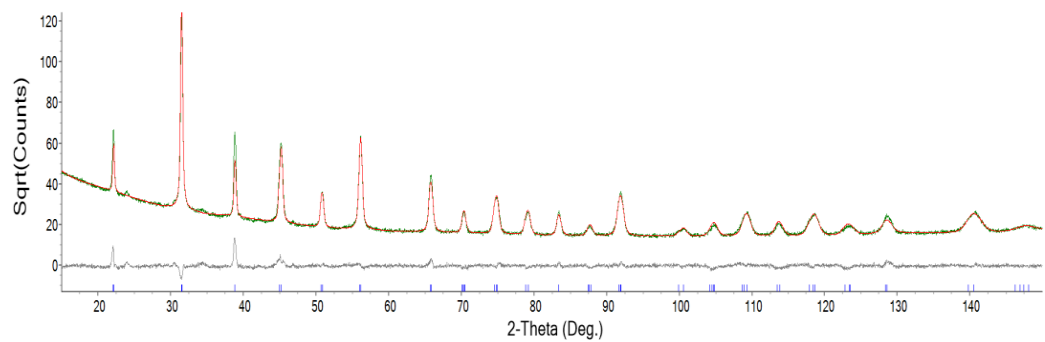

## (c) 10 nm washed

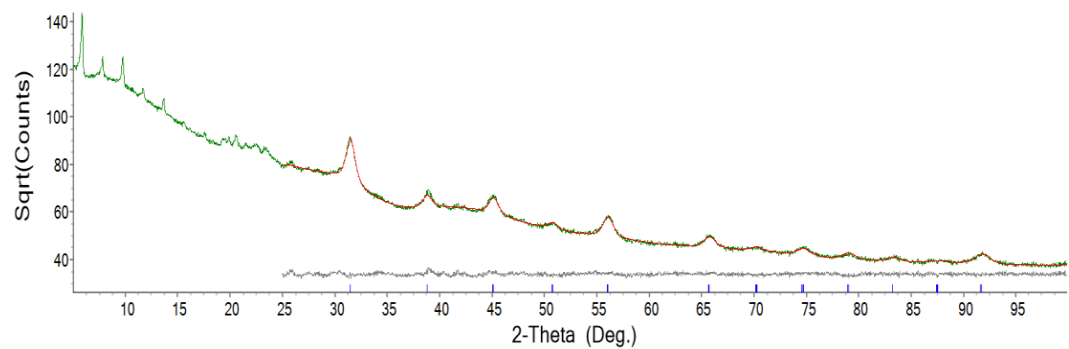

**Fig. S1**. Rietveld fits of the tetragonal P4mm structure for the (a) 700 nm, (b) 50 nm, and (c) 10 nm washed $BaTiO_3$ samples. The 700 nm sample was used as a bulk reference and the Scherrer model [4] was used to fit the data. For the 10 nm washed $BaTiO_3$ sample, only the region above 25° was used to avoid the major peaks from the barium oleate component. In the figures, the red curve is the model fit, the green curve is the experimental data, and the lowest curve (black) is the difference between the experimental and model data; the blue vertical inner ticks on the x-axis are the positions of the model peaks (hkl). Note the prominent barium oleate peaks at small angles in (c). We use the simple tetragonal structure to model the system in order to reduce the number of free parameters. The true sample may be orthorhombic. We note that the 10 nm sample is not cubic (a is not equal to c). Label 700 nm-Bulk is abbreviated as 700 nm in the main text.

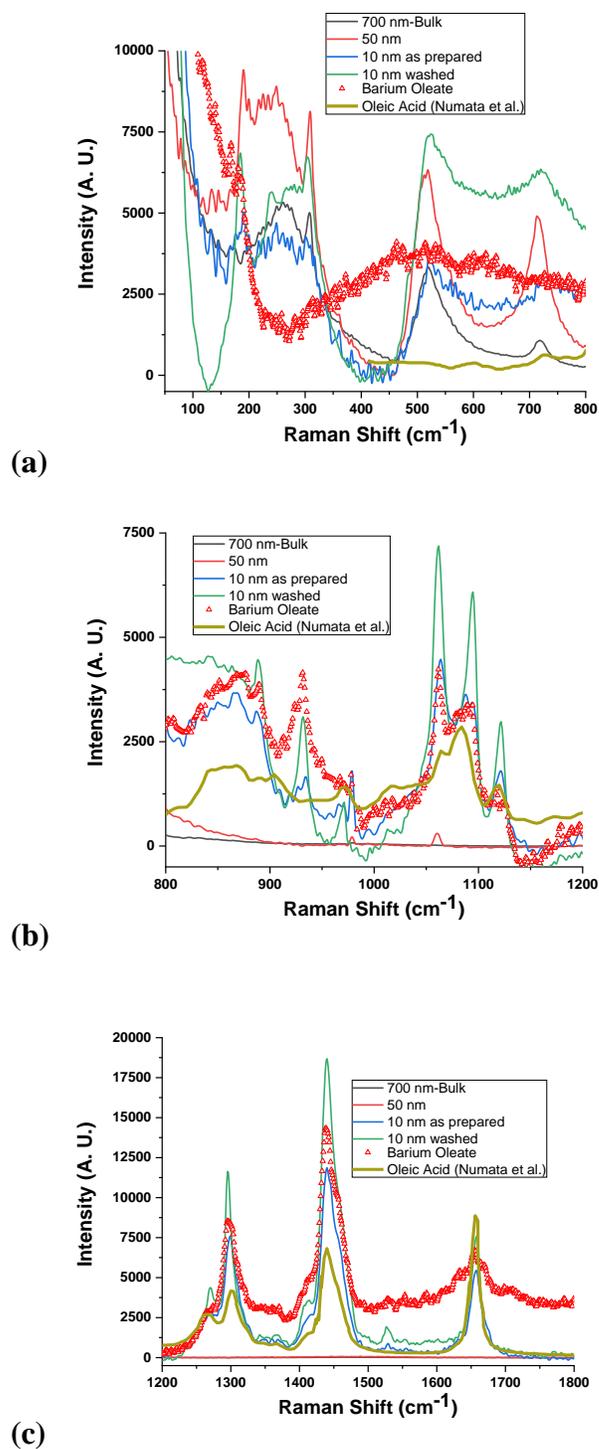

**Fig. S2.** Expanded Raman spectra based on Fig. 1 showing 700 nm and 50 nm $BaTiO_3$ reference samples, 10 nm nanoparticle $BaTiO_3$ (washed and as-prepared), oleic acid (taken from Ref. [5]), and barium oleate. The data cover the ranges: (a) 50 to 800 $cm^{-1}$, (b) 800 to 1200 $cm^{-1}$ and (c) 1200 to 1800 $cm^{-1}$. The barium oleate spectra matches well with the nanoparticle measurements for the full spectral range except that the nanoparticle spectra are significantly more narrow due to high crystallinity. A. U. indicates arbitrary units. Label 700 nm-Bulk is abbreviated as 700 nm in the main text.

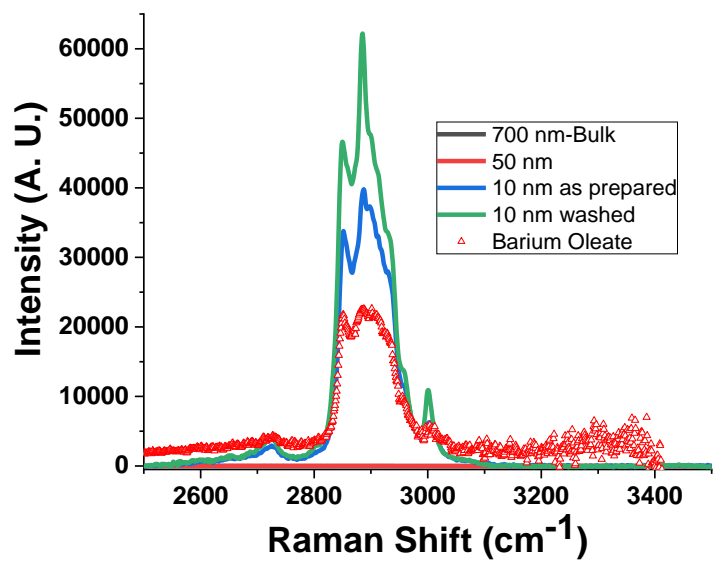

**Fig. S3.** High energy region of the Raman spectrum corresponding to the data in Fig. S2. Again, note that the spectrum of the 10 nm samples is similar to that of barium oleate, except that the 10 nm samples have sharper features due to high structural order. Label 700 nm-Bulk is abbreviated as 700 nm in the main text.